\documentstyle[psfig]{caosp}
\begin{document}
\pubyear{1998}
\volume{999}
\firstpage{8}
\hauthor{A. Hui-Bon-Hoa}
\title{Abundances of metals in five nearby open clusters}
\author{A. Hui-Bon-Hoa}
\institute{D.A.E.C., Observatoire de Meudon, Meudon F-92195, France}
\date{December 30, 1997}
\maketitle
\begin{abstract}
Abundances of Mg, Ca, Sc, Cr, Fe, and Ni are derived for A stars of five
nearby open
clusters of various ages using high resolution spectroscopy. We point out a
correlation
between the abundance of Ca and that of Sc, suggesting that the abundance
anomalies of these
elements arise from the same physical process. Pronounced Am patterns are
rather found in
the oldest cluster stars whereas younger targets show weaker Am anomalies
and atypical
patterns for some of them.
\end{abstract}
\section{Introduction}
The abundance anomalies of the Am stars (and, as a rule, of most of the
Chemically Peculiar
stars) are nowadays broadly considered to be produced by microscopic
diffusion. Indeed, the
observed anomalies are consistent with the computed diffusion velocities
for many elements.
Nevertheless, the detailed understanding of the Am phenomenon requires more
thorough
studies since the stratification process is time-dependent and affected by
the other physical
processes at play in the stellar medium. Calcium and scandium are elements
of special
interest since their deficiencies are usually used to detect Am stars
(Conti 1970; Preston
1974). The evolution of the abundances of Ca and Sc in the superficial
layers of an A star has
been computed by Alecian (1996) assuming no helium convection zone, as
suggested by the
diffusion model for Am stars. His results show that the behaviour of both
elements are
strongly dependent on the strength of the large scale motions introduced in
the computations,
namely, a mass-loss and the extension of the superficial mixing zone
beneath the convective
zone. In some cases, phases of overabundance occur when the star is around
10 million years
old. So, the classification criterion of Am stars based on the calcium or
scandium deficiency
is questionable for young A stars.

Abundance determinations in young main-sequence A stars can constrain such
computations.
Open cluster stars are the best candidates since their age is known with
much greater
accuracy than for field stars. Few abundance studies in open clusters have
been done up to
now (see Burkhart \& Coupry 1997 and references therein). The use of
electronic detectors
(Reticon and CCDs) and the subsequent improvement in the quality of the
spectra made
possible the study of the lithium abundance and, by the way, renewed the
interest for open
clusters: observations of A-type stars were carried out by Boesgaard (1987)
in Coma, and
by Burkhart \& Coupry (1989; 1997) in the Hyades and Pleiades.

This paper summarizes the work undertaken in collaboration with G. Alecian
and C. Burkhart
concerning the abundance of calcium and scandium in open cluster A stars.
More details are
provided in Hui-Bon-Hoa et al. (1997) and Hui-Bon-Hoa \& Alecian (1998).
\section{Observational data}
\subsection{The sample}
We chose 34 A-type stars belonging to the northern open clusters:
$\alpha$ Per, Pleiades,
Coma, Hyades, and Praesepe (see Table 1). The ages of these clusters are
well spread over
the time interval covered by the simulations of Alecian (1996). The
selection criteria for
each star were:
\begin{list}{}{}
\item - rotational velocity compatible with abundance determinations using
the lines of our
wavelength domain;
\item - availability of \textit{uvby $\beta$} photometry for the estimation
of effective
temperature and surface gravity.
\end{list}
We have included all A stars that match these criteria in our sample since
Alecian (1996)
suggests that a phase of overabundance of Ca and Sc can occur for young Am
stars.

Three field stars have also been observed: HR 178 (HD 3883), Sirius
($\alpha$ CMa, HD
48915), and Procyon ($\alpha$ CMi, HD 61421).
\begin{table}[t]
\small
\begin{center}
\caption{Basic data for the clusters.}
\begin{tabular}{ccc}
\hline\hline
Cluster&Age (yr)&Number of stars\\
\hline
$\alpha$ Per&$\mathrm{5\,10^{7}} (1)$&6\\
Pleiades&$\mathrm{10^{8}} (1)$&9\\
Coma Ber&$\mathrm{4.3\,10^{8}} (2)$&6\\
Hyades&$\mathrm{6.7\,10^{8}} (2)$&6\\
Praesepe&$\mathrm{7.6\,10^{8}} (2)$&7\\
\hline\hline
\end{tabular}
\begin{list}{}{}
\item References: (1) Meynet et al. (1993); (2) Boesgaard (1989)
\end{list}
\end{center}
\end{table}
\subsection{Instrumentation}
The spectra were obtained during two runs in December 1995 and December
1996 at the
Observatoire de Haute-Provence using the AURELIE spectrograph at the
coud\'{e} focus of
the 152~cm telescope. This instrument was set to yield a spectral
resolution of about 34000
(linear dispersion of 5~\AA$\mathrm{mm^{-1}}$). The limiting \textit{V}
magnitude was
around 9 for an integration time of 3 hours and a typical signal-to-noise
ratio of 300. We
chose the spectral interval 5495--5620~\AA\,, which allows abundance
determinations of
Mg, Ca, Sc, Cr, Fe, and Ni.
\section{Abundance determination}
We used model atmospheres computed with the ATLAS9 code (Kurucz 1992a, b). The
atmospheric parameters are derived from \textit{uvby $\beta$} photometry
through the
grids of Moon \& Dworetsky (1985) which are still quite reliable for single
stars as shown
by Napiwotski et al. (1993).

The abundance analysis is carried out assuming LTE with a series of codes
written by M.
Spite (1967, 1996 private communication). We used solar oscillator
strengths deduced from
a solar spectrum obtained with the same instrumentation as for the
programme stars. We
mostly used a curve of growth method and the equivalent widths are measured
with a
procedure developed by Cayrel et al. (1985), which fits gaussian profiles
to the observed
spectrum. When the rotational velocity is above 30~km/s, the gaussian
profile is not
suitable anymore and the lines are fitted with a rotational profile.
\begin{figure}[ht]
\psfig{figure=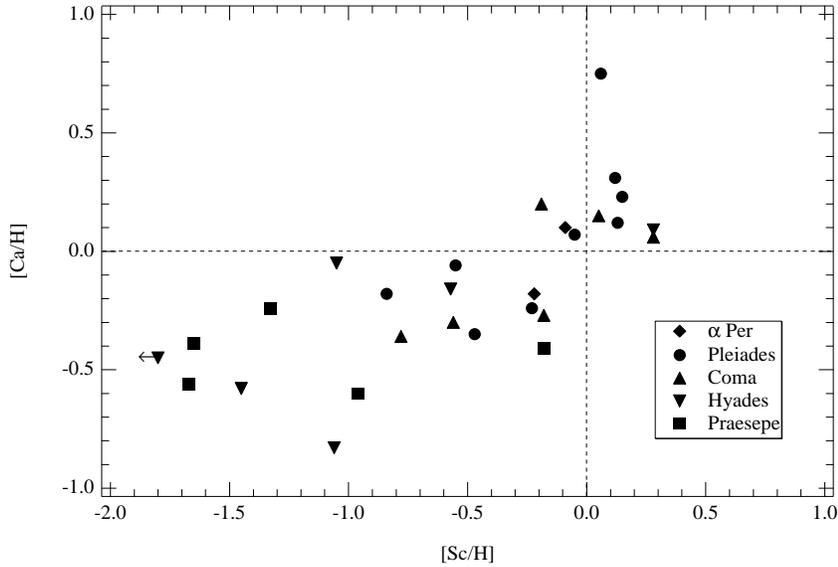,width=11.5cm}
\caption{[Ca/H] vs. [Sc/H] for the cluster stars. The arrow means an upper
limit for the Sc
abundance.}
\end{figure}
\section{Results and discussion}
\begin{figure}[ht]
\psfig{figure=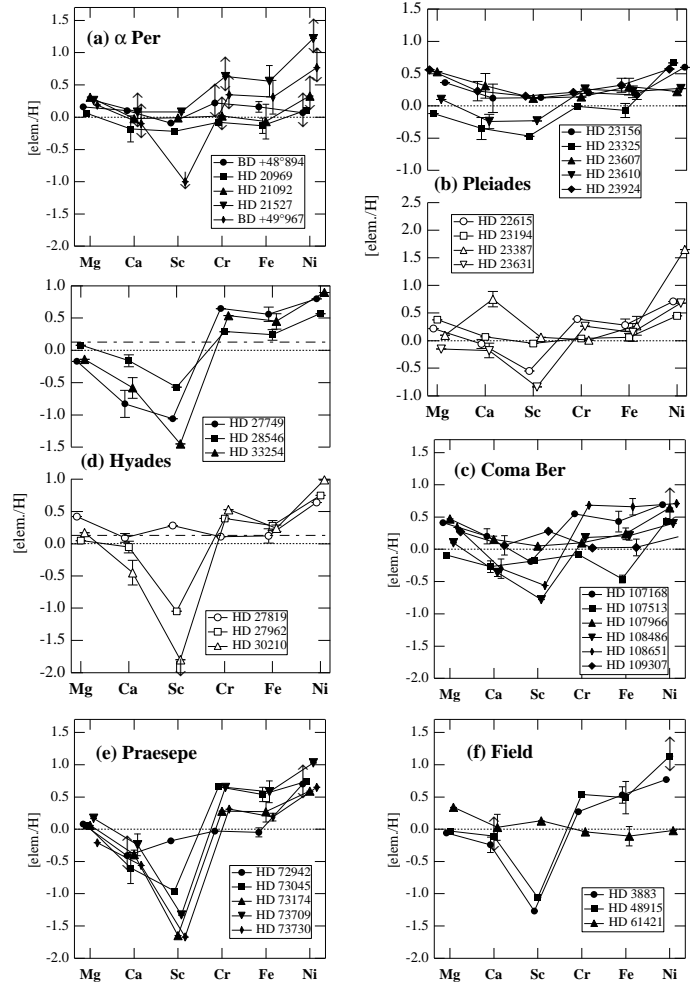,width=10cm}
\caption{Abundance patterns for the stars of: \textbf{a} $\alpha$ Per;
\textbf{b} the
Pleiades; \textbf{c} Coma; \textbf{d} the Hyades; \textbf{e} Praesepe;
\textbf{f} field
stars. The metallicity of the Hyades is indicated by
a dash-dotted line.}
\end{figure}
As usual, the script [X] for any quantity X means
$\mathrm{log\,(X)_{*}\,-\,log\,(X)_{\sun}}$. In Fig.~1, we can see a loose
correlation
between [Ca/H] and [Sc/H] for our cluster stars. This suggests that the
anomalies of these
two elements come from the same physical process. Besides, the more pronounced
deficiencies (left part of the graph) are found in the oldest stars
(members of the Hyades and
Praesepe). Younger targets (in $\alpha$ Per, Pleiades) show weak
underabundances or
marginal overabundances (right part of the graph).

The abundance patterns for our sample stars are shown in Fig.~2. In
ordinates are the
logarithmic differences between the abundance value in the star of concern
and the solar one
for Mg, Ca, Sc, Cr, Fe, and Ni. The corresponding points are linked for
each star. An arrow
means upper limit and error bars ended by arrows denote very uncertain
values. We can see
that the Am pattern is well-marked in the oldest cluster stars (Hyades and
Praesepe) of our
sample as well as for the field Am stars. In the youngest clusters
($\alpha$ Per, Pleiades),
the Am pattern is almost absent and several objects show atypical patterns
with marginal
overabundances of Ca and/or Sc. This would suggest that stars of these
clusters are in
transient phases of the stratification process.

The marginal overabundances of Ca and Sc in $\alpha$ Per and Pleiades are
not strong
enough to confirm the phase of overabundance predicted by Alecian (1996).
Either the
youngest clusters of our sample are already too old and their stars have passed
the phase of overabundance or this phase does not exist. In this last case, the
extension of the mixing
zone should be less than one pressure scale height and the mass-loss rate about
$\mathrm{10^{-14}M\sun/yr}$.

These conclusions need to be confirmed by observations of more clusters and
younger
targets. Also, elements that could help a better understanding of the Am
phenomenon
deserve to be studied (rare earths for instance).

\acknowledgements
Many thanks to M. Spite who has kindly provided us with her codes and to C.
van't Veer for
having introduced us to the ATLAS9 code. Thanks also to G. Michaud and S.
J. Adelman for
useful suggestions and to the OHP staff for their excellent support.

\end{document}